\documentclass[prc,twocolumn,showpacs,showkeys,nofootinbib,preprintnumbers,amsmath,amssymb]{revtex4}

\usepackage{graphicx}
\usepackage{dcolumn}
\usepackage{bm}

\begin{document}

\title{Pseudorapidity shape of elliptic flow as signature for 
fast equilibration in relativistic heavy-ion collisions at energies 
up to $\sqrt{s_{NN}} = 200~{\rm GeV}$}

\author{J. Bleibel}
\author{G. Burau}
\author{Amand Faessler}
\author{C. Fuchs}

\affiliation{
Institute for Theoretical Physics, University of T\"ubingen, 
Auf der Morgenstelle 14, D-72076 T\"ubingen, Germany
}


\begin{abstract}
The implications of parton recombination processes on the dynamics of 
ultrarelativistic heavy-ion reactions are investigated. To do so, 
the quark-gluon string transport model has been extended for partonic 
recombination and fusion processes. Parton recombination leads to short 
equilibration times and improves significantly on the theoretical 
description of measured directed and elliptic flow, i.e., $v_1$ and $v_2$, 
distributions in Au+Au collisions at $\sqrt{s_{NN}} = 200~{\rm GeV}$, 
in particular what concerns their pseudorapidity dependence. The shape 
of $v_2(\eta)$ is found to be closely related to fast thermalization.
\end{abstract}



\maketitle


\section{Introduction}
\label{intro}

Among the various experimental studies of ultrarelativistic heavy-ion 
collisions operates the Relativistic Heavy Ion Collider (RHIC), at 
Brookhaven National Laboratory since 2000, to investigate gold-on-gold 
collisions up to $\sqrt{s_{NN}} = 200~{\rm GeV}$. After many years of 
operation strong experimental evidence has been accumulated that at RHIC 
energies indeed a new state of matter is created that is qualitatively 
different from a hadron gas (see Ref. \cite{QM2005} and references therein). 
This state seems, however, not to behave like a weakly interacting gas of 
deconfined partons -- as could have been naively expected -- but rather like 
a strongly coupled quark-gluon plasma (sQGP). One argument toward such a 
scenario is the large elliptic flow observed at RHIC 
\cite{Ackermann:2000tr,Park:2001gm,Manly:2002uq,Back:2004mh,Adler:2003kt}. 
The development of strong elliptic flow requires short equilibration
times and large pressure gradients to drive the dynamics, both being
characteristic features of a strongly interacting system 
\cite{Heinz:2001xi,Shuryak:2003xe}.

Another evidence for this picture is provided by the hadron species dependence 
of the elliptic flow \cite{Adler:2003kt,Adams:2003zg}. The observed scaling 
with the number of constituent quarks can naturally be explained by the 
assumption that the elliptic flow is to most extent already created in the 
partonic phase and transferred to the hadrons through a partonic recombination 
mechanism \cite{Dover:1991zn,Hwa:2002tu,Hwa:2005ay,Greco:2003xt,Greco:2005jk,Molnar:2003ff,Molnar:2005wf,Fries:2003kq}.

In the present work we demonstrate that not only the size but also the 
shape of the observed pseudorapidity profiles of the elliptic flow $v_2$ 
provide strong evidence for a partonic recombination mechanism in 
combination with fast thermal equilibration inside a sQGP phase. 
As basis for these investigations serves a microscopic quark-gluon string 
model (QGSM) that has been extended to allow for a partonic recombination 
procedure motivated by parton coalescence models \cite{Greco:2003xt,Greco:2005jk,Molnar:2003ff,Molnar:2005wf}, i.e., to model effectively the dynamics of a 
strongly coupled quark plasma formed in the very dense stages of 
ultrarelativistic heavy-ion collision. By construction, string cascade models 
do not contain an explicit quark-hadron phase transition. However, during the 
temporal evolution of a heavy-ion reaction a dense and strongly interacting 
plasma is formed within such approaches as well. The system consists of 
partons and color-flux tubes (or strings). Thus, such models can serve as 
starting base to study the dynamics of the dense and strongly interacting 
medium created at RHIC, in particular because transverse as well as elliptic 
flow at SPS energies are well reproduced within the string-cascade
approach \cite{DF_prc00,DF_prc01,LPX99,Petersen:2006vm}.

It was shown in Refs. \cite{Burau:2004ev,Zabrodin:2005pd} that the standard 
version of the microscopic quark-gluon string cascade model (QGSM) is able 
to describe fairly well the bulk properties of the elliptic flow 
$v_2$ measured in $\sqrt{s_{NN}}=200~{\rm GeV}$ Au+Au reactions at RHIC. 
However, a difference in shape in particular between the pseudorapidity 
distributions of the elliptic flow at midrapidity determined in the experiment 
and in the simulation has rankled. The shape of the $v_2$ distributions 
have been found to be closely related to anisotropies in the corresponding 
energy density profiles and the degree of equilibration \cite{Bleibel:2005gx}. 
Although the system reaches quickly some sort of a pre-equilibrium stage 
within standard QGSM the processes included are not sufficient to drive the 
system to a completely thermalized state. As shown in the following, this 
changes dramatically when partonic recombination processes are included. 
The final consequences for the anisotropic flow turned out to be even more 
remarkable.


\section{Quark-gluon string model with parton recombination}
\label{mcqgsm}

The  \textsl{standard} Monte Carlo version of the quark-gluon string 
model \cite{QGSM0} that serves as the basis for the present investigations 
has been described in detail in Refs. \cite{QGSM1c,Burau:2004ev}. 
Based on Gribov-Regge Theory, the production of hadrons is described by 
excitation and decay of open strings with different quarks or diquarks on 
their ends. In that sense the model already incorporates the partonic 
structure of hadrons and therefore can provide a framework for the inclusion 
of partonic recombination processes. To trigger these processes, a certain 
critical (energy) density is needed to allow the hadrons to overlap. 
Concerning the implementation of partonic recombination processes we apply 
essentially the method proposed in Ref. \cite{Greco:2003xt}: 
The partons of three hadrons are allowed to enter into a recombination 
process if their spatial distance, in the center-of-mass (c.m.) frame of 
the corresponding hadrons, is less than $\Delta_x = 0.85~{\rm fm}$. 
From all possible triplets of hadrons satisfying these spatial constraints 
one is randomly chosen according to usual Monte Carlo methods and the 
corresponding hadrons are then decomposed into their constituent partons 
(quark-antiquark or quark-diquark). 
Each parton is given a momentum fraction $z$ of the initial momentum of its 
hadron and, additionally, the partons of each pair obtain a transverse 
momentum $p_T$ of opposite sign. Both, $z$ and $p_T$ are generated from the 
standard parton distribution functions used within QGSM. 
The (random) selection of a triplet of hadrons by the coordinate space 
criterion $\Delta_x$ is based on the approximation that distances of hadrons 
are equal to (averaged) distances of their constituent partons. In that sense 
the selection procedure really acts on the partonic level.

The recombination process itself requires also an overlap of the 
participating partons in momentum space. The distance in momentum space 
between the pairs of partons is evaluated in the center-of-mass frame. 
The probability for recombination of two partons is then given by the 
covariant distribution \cite{Greco:2003xt}
\begin{eqnarray}
  f_2(x_1,x_2,p_1,p_2)&=&\frac{9\,\pi}{2\Delta_x^3\Delta_p^3}\,
  \Theta\left[\Delta_x^2-(x_1-x_2)_{\rm c.m.}^2\right]\nonumber\\
  &&\hspace{-1.8cm}\times\,\Theta\left[\Delta_p^2-(p_1-p_2)_{\rm c.m.}^2\right]~.
\label{probdist2}
\end{eqnarray}
Additionally, if the partons of three mesons participate, recombination of 
three quarks and three antiquarks is possible with a probability 
distribution \cite{Greco:2003xt}
\begin{eqnarray}
  f_3(x_1,x_2,p_1,p_2)&=&\frac{9\,\pi}{2\Delta_x^3\Delta_p^3}\,
  \Theta\left[\Delta_x^2-(x_1-x_2)_{\rm c.m.}^2\right]\nonumber\\
  &&\hspace{-1.8cm}\times\,\Theta\left[\Delta_p^2-(p_1-p_2)_{\rm c.m.}^2\right]\nonumber\\
  &&\hspace{-1.8cm}\times\,\frac{9\,\pi}{2\Delta_x^3\Delta_p^3}\,
  \Theta\left[\Delta_x^2-(\frac{x_1+x_2}{2}-x_3)_{\rm c.m.}^2\right]\nonumber\\
  &&\hspace{-1.8cm}\times\,\Theta\left[\Delta_p^2-(p_1+p_2-2\,p_3)_{\rm c.m.}^2\right]~,
\label{probdist3}
\end{eqnarray}
leading effectively to a fusion of three mesons into a baryon-antibaryon pair. 
We want to stress at this point that our fusion process, which allows 
pseudoscalar and vector mesons to ``rearrange'' to $B\bar{B}$ on the partonic 
level (no cross sections have been introduced), is qualitatively consistent 
with the method applied in Ref. \cite{Cassing:2001ds}, where a quark 
rearrangement model for $B\bar{B}$ annihilation to three mesons (pseudoscalar 
and vector, which later decay to additional pions) has been employed. 
For example: The fusion of two $\rho$ mesons and a pion into $p\bar{p}$ is 
kinematically favored compared to the fusion of three pions (phase space). 
However, a more quantitative investigation has to be delayed to future work.

The momentum coalescence radius $\Delta_p$, which enters in the distributions 
(\ref{probdist2}) and (\ref{probdist3}), is given by the uncertainty 
principle $\Delta_x \Delta_p = 1$, where the actual distance between the 
participating partons is used for $\Delta_x$. To exclude recombination of 
particles with highly different momenta high-momentum cutoffs of 
$\Delta_{p,{\rm max}}=m_1+m_2$ for recombination processes $3H \rightarrow 3H$ 
($H=M,B,\bar{B}$) and $\Delta_{p,{\rm max}}=m_1+m_2+m_3+0.2~{\rm GeV}$ for the 
aforementioned fusion process $3M \rightarrow B\bar{B}$ are used, where $m_i$ 
corresponds to typical constituent masses. 
We want to stress that recombination and fusion happen on the partonic 
level, i.e., the notation denotes the contained partons, which participate 
in those processes, in the following way: $M = (q\bar{q})_M$, $B = (qqq)_B$, 
and $\bar{B} = (\bar{q}\bar{q}\bar{q})_{\bar{B}}$, respectively.

The recombination (or fusion) processes can take place if the quantum 
numbers allow the partons of at least two of the three overlapping 
hadrons to recombine into new hadronic correlations, whereas this process 
is mediated by the partons of the third hadron. 
Effectively, this is a three-body interaction or at most a kind of in-medium 
two-body interaction but no additional two-body interaction in the vacuum. 
Hence the total vacuum hadron-hadron cross section implemented in the QGSM 
is not changed. 
Anyway, as mentioned above no explicit in-medium cross sections 
have been introduced because all recombination processes are dynamically 
treated on the partonic level. Nevertheless it is possible to estimate 
effective cross sections for the implemented partonic recombinations 
between the hadronic correlations. The maximum value of the inclusive 
cross section for recombination processes like $3M \rightarrow 3M$ or 
$3M \rightarrow B\bar{B}$ is basically determined by the aformentioned 
spatial coalescence radius $\Delta_x$. The allowed maximum value of 
$0.85~{\rm fm}$ \cite{Greco:2003xt} is geometrically related to a cross 
section not larger than about $23~{\rm mb}$. Because the dense medium 
produced in the overlap zone of a highly energetic Au+Au collision is 
dominated by ``pionic correlations'', the recombination of partons coming 
from and going to two ``pions'' in presence of the partons of a third 
(medium) ``pion'' happens most frequently. Nearly 90\% of all recombination 
processes with three pionic states in the initial channel are reactions of 
this kind, whereas the ``fusion process'' $3\pi \rightarrow p\bar{p}$ is 
distinctly suppressed. Accordingly, the effective cross section of the 
latter process is several orders of magnitude smaller than the estimated 
cross section of the first process, which turns out to be not larger than 
$20~{\rm mb}$. 
Furthermore, we have estimated effective cross sections for the 
partons (quarks and antiquarks) involved in recombinations by their distance 
in momentum space using $\sigma_{q\bar{q}} \propto (\Delta_p(q,\bar{q}))^{-2}$. 
Depending on the center-of-mass energy of the (anti)quarks, the partonic 
cross section is in the order of a few millibarn but definitely not larger 
than the aforementioned $23~{\rm mb}$. The rate of recombination processes 
is determined by the cutoff parameter for high momenta $\Delta_{p,max}$: 
increasing $\Delta_{p,max}$ increases the recombination rate. However, the 
additional recombination processes with highly different momenta have only 
very small cross sections. Because the bulk of recombination processes 
happens to take place with $\Delta_p(q,\bar{q})$ close to the cutoff 
$\Delta_{p,max}$, an averaged cross section would yield a value comparable 
to parton scattering cross sections used, e.g., in the AMPT model 
\cite{Chen:2004vh}.

The implemented procedure always ensures that only physical particles can be 
final hadronic states, e.g., no final diquark states are possible. Also the 
reaction $HB\bar{B} \rightarrow H3M$, i.e., the partons of a hadron and a 
baryon-antibaryon pair form finally a hadron and three mesons, which contains 
as a subprocess the backreaction to ``meson fusion'', is not included in the 
recombination scheme because annihilation of baryons and antibaryons is 
already implemented in the QGSM as a standard two-particle reaction 
$B+\bar{B} \rightarrow X$. Furthermore, the model allows for the formation 
of resonance states and their decay. Consequently, subprocesses such as 
$2M \rightarrow 3M$, which can also increase the particle number, are 
incorporated in the model. 
However, if the outcome of a recombination process would yield any unphysical 
particle it cannot take place. Either an other valid recombination process 
is then chosen or, if none is possible, no recombination occurs.

Apart from recombining to new mesons, (anti)baryons or baryon-antibaryon 
pairs, quark-antiquark annihilation is possible when the partons of three 
mesons participate in a recombination process. Thus a quark-antiquark pair 
of the same flavour, but belonging to different mesons, may  annihilate with 
a redistribution of its energy and momentum to the new mesons formed by the 
remaining partons. The probability of this annihilation process 
$3M \rightarrow 2M$ with respect to $3M \rightarrow 3M$, i.e., 
$P_{\rm a} = 0.04$, has been adjusted to reproduce the experimental 
$dN/d\eta$ charged hadron multiplicities. By means of this annihilation 
process $3M \rightarrow 2M$ an effective backward reaction for diffractive 
scattering is included. 
From the possible recombination and annihilation processes, including the case 
that nothing happens at all, i.e., all partons recombine to the original 
hadrons, the actual reaction is randomly chosen. These processes are checked 
for all combinations of overlapping hadrons, making thereby sure, however, 
that the particular partons of the selected hadron triplet can only once per 
time step -- which is about $10^{-5} \dots 10^{-3}~{\rm fm/c}$ -- participate 
in such processes. 
If no recombination (annihilation) or fusion processes take place normal 
elastic or inelastic scattering occurs.

Charge conservation is automatically guaranteed within our approach, but 
conserving energy and three-momentum simultaneously is not possible for 
the recombination processes described above. Generally, there are problems 
to conserve simultaneously energy and momentum within partonic 
coalescence/recombination approaches, see, e.g., Ref. \cite{Fries:2003kq}, 
where conservation of momentum has been chosen whereas conservation of energy 
has been violated. In our model, conservation of energy is violated in most 
recombination processes on the level of few percentages only. 
However, applying a rescaling procedure for the momenta in the center-of-mass 
system of the produced hadronic correlations, we are able to ensure 
simultaneous conservation of energy and three-momentum with a precision 
better than 1 MeV. To conserve the initial energy the three-momentum 
components of all particles involved in a particular reaction are scaled by a 
constant factor which is determined iteratively. Moreover, by such a procedure 
ratios of momenta, e.g., anisotropic flow coefficients, are remain unchanged. 
In other words, the anisotropic flow is not artificially influenced by the 
momentum rescaling.

All particles are allowed to interact via the recombination procedure 
described above, even non-formed (pre-) hadrons. 
In this spirit the model does not create a ``system of free partons'' but 
effectively emulates a medium of very strongly correlated partons, i.e., 
quark-antiquark and (anti)quark-(anti)diquark states.
To not shorten or modify the formation times of the nonformed ``hadrons'', 
their formation time is now interpreted on the quark level, meaning the time 
for the constituent partons to be fully created. As a consequence of that 
each hadron gets a formation time for each constituent [quark, antiquark or 
(anti-) diquark] $t^{\rm f}_1$ and $t^{\rm f}_2$. Within a normal production 
process such as string-breaking $t^{\rm f}_1$ and $t^{\rm f}_2$ are equal. 
For the recombination mechanism the newly produced hadronic states get the 
formation times of their constituents. The new hadron will be fully formed 
if the lifetime exceeds the larger formation time and partially formed 
(like a leading hadron) if the lifetime exceeds the smaller formation time. 
These partially formed hadrons are allowed to rescatter assuming additive 
constituent cross sections.


\section{Equilibration and anisotropic flow of charged hadrons}
\label{flow}

\begin{figure}[ht!]
\includegraphics[scale=0.5235]{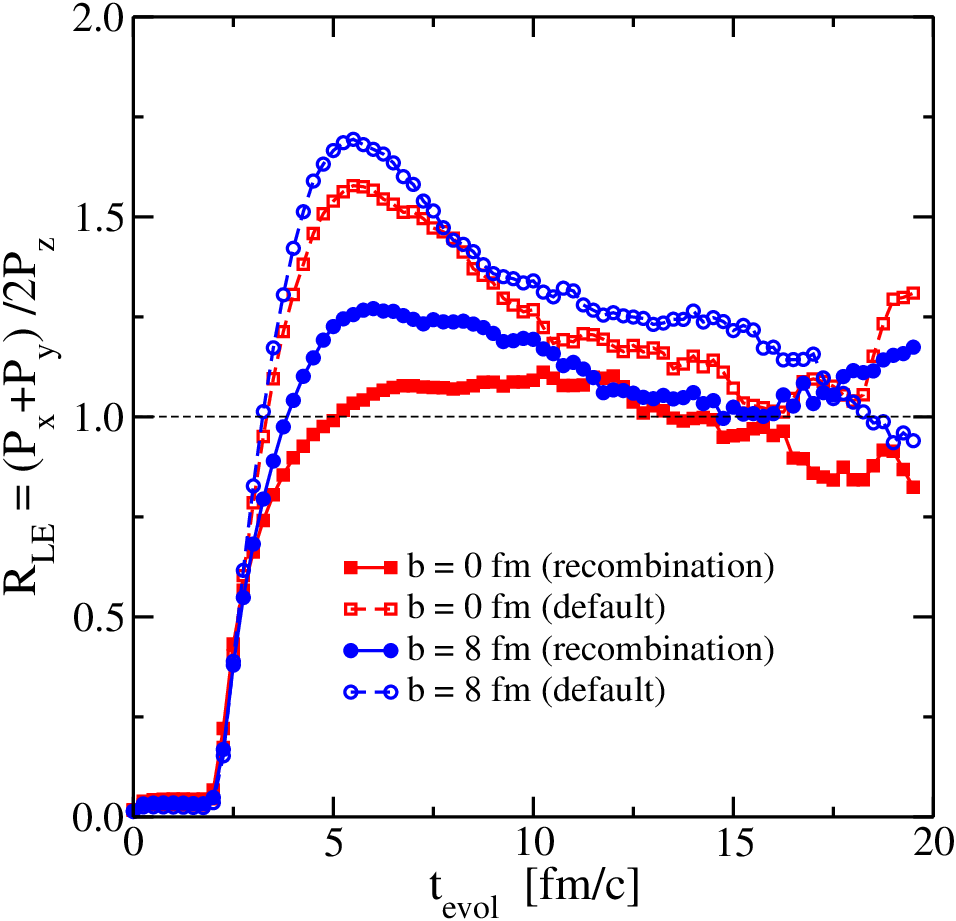}
\caption{Influence of parton recombination on the kinetic equilibration time. 
The figure shows the time dependence of the equilibration ratio $R_{\rm LE}$ 
in the central cell of the overlap zone for central and semiperipheral 
Au+Au reactions at $\sqrt{s_{NN}} = 200~{\rm GeV}$ for QGSM simulations 
with (filled symbols) and without (open symbols) partonic recombination 
processes.
\label{EQ1}}
\end{figure}

First we examine the influence of parton recombination on the kinetic 
equilibration in the central cell of the overlap zone of Au+Au 
collisions. This cell is given by $-2~{\rm fm} < x, y, z < 2~{\rm fm}$ for 
a central Au+Au reaction with impact parameter $b=0~{\rm fm}$ and 
$2~{\rm fm} < x < 6~{\rm fm}$, $-2~{\rm fm} < y, z < 2~{\rm fm}$ for a 
semiperipheral reaction ($b=8~{\rm fm}$), respectively. The corresponding 
equilibration ratio $R_{\rm LE} = (P_x + P_y)/2P_z$ is determined by the 
pressure components in x, y and z directions, $P_{x,y,z}$ (see discussion 
in Ref. \cite{Bleibel:2005gx}). The QGSM results with and without 
implementation of parton recombination are depicted in Fig. \ref{EQ1}. 
A difference in the degree of local kinetic equilibration is clearly seen 
when one compares the standard QGSM results with the parton recombination 
scenario. 
The latter mechanism leads very effectively to a much faster and smoother 
equilibration for both, the central and semiperipheral collisions. 
Essentially, the decomposition and recombination procedure reshuffles rapidly 
the momenta of partons and finally of the hadrons which drive the system 
undoubtedly into kinetic equilibrium within short times. 
For central reactions, the matter in the central cell is practically fully 
equilibrated, i.e., $R_{\rm LE} \approx 1$, after very few $\rm fm/c$ time of 
evolution. Here, of course, the effect of the recombination mechanism is 
strongest, because it is clearly a density-dependent mechanism. But even for 
semiperipheral collisions the local pre-equilibrium stage identified by 
$R_{\rm LE} > 1$ is remarkably shortened compared to the scenario without 
recombination of partons in the dense stages of the expanding medium.
It should be noticed that the offset of $t_{\rm evol}$ seen in Fig. \ref{EQ1} 
is due to the chosen cell size and is of technical nature. 
In the first few time steps the pressure is dominated by the longitudinal 
flow of the nuclei penetrating the cell. Disregarding this offset one can 
see that parton recombination reduces the equilibration time of the system 
approximately by a factor of 5 from $\sim 10~{\rm fm/c}$ to 
$\sim 2~{\rm fm/c}$.

The interplay between fast thermal equilibration and the amount of 
anisotropic flow, in particular the elliptic flow, measured in high-energy 
heavy-ion reactions is strongly debated (see e.g. \cite{Bhalerao:2005mm,
Ackermann:2000tr,Park:2001gm,Manly:2002uq,Back:2004mh,Adler:2003kt,
Adams:2003zg,BeltTonjes:2004jw} and references therein). 
In Ref. \cite{Bleibel:2005gx} the idea has been supported that a fast and 
complete thermal equilibration is not strictly necessary to produce 
\textsl{large} elliptic flow. Nevertheless, a conspicuous difference in shape 
in particular between the pseudorapidity distributions of the elliptic flow 
at midrapidity determined in the experiment and in the simulation was 
observed. Hence it is very natural to study the effect of the quark 
recombination mechanism on the azimuthal anisotropy parameter $v_2$ 
within the QGSM \cite{Burau:2004ev}. 
\begin{figure}[b]
\includegraphics[scale=0.525]{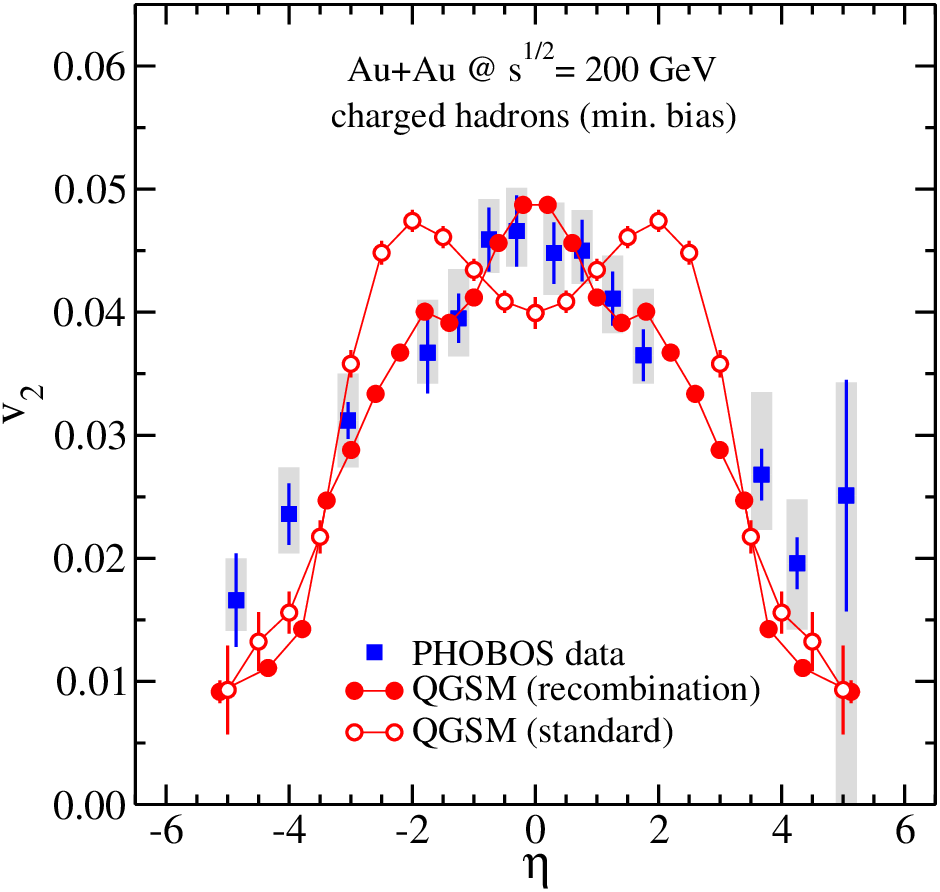}
\caption{Influence of parton recombination on the pseudorapidity dependence 
of the elliptic flow. Results for $v_2(\eta)$ of charged hadrons from the 
standard QGSM (open symbols) \cite{Burau:2004ev} and those obtained with 
parton recombination (filled symbols) are compared to  PHOBOS data of minimum 
bias Au+Au collisions at $\sqrt{s_{NN}} = 200~{\rm GeV}$ \cite{Back:2004mh}. 
The statistical error bars are shown together with the systematic errors of 
the experimental data (gray boxes).
\label{fig:v2etahchar1}}
\end{figure}
Figure \ref{fig:v2etahchar1} shows the pseudorapidity dependence of the 
elliptic flow $v_2$ of charged hadrons for minimum bias Au+Au collisions 
at $\sqrt{s_{NN}} = 200~{\rm GeV}$. 
The QGSM results including parton recombination and the standard approach 
without recombination \cite{Burau:2004ev} are compared with the experimental 
data of the PHOBOS Collaboration \cite{Back:2004mh}. 
The elliptic flow obtained within the standard QGSM displays a strong 
in-plane alignment in accordance with the experimental findings. 
At midrapdity $|\eta| < 1$ the flow parameter $v_2$ is almost constant, 
but then it rises up slightly followed by a rapid drop at $|\eta| > 2$. 
The emergence of this peculiar double bump structure in $v_2(\eta)$ was 
strongly connected with the model dynamics (for more details, see 
Ref. \cite{Burau:2004ev}). In contrast, the experimentally observed elliptic 
flow shows a pronounced peak at midrapidity and a steady decrease for 
$|\eta| > 1$ \cite{Back:2004mh,Back:2004je}. This behaviour is  remarkably 
well reproduced when partonic recombination processes are taken into account. 
The parton rearrangement processes in the dense medium lead to a 
redistribution of the elliptic flow of the final hadrons toward midrapidity, 
i.e., $v_2$ is accumulated at $|\eta| \approx 0$, whereas it is distinctly 
reduced in the region around $|\eta| \approx 2$. Thus, the double bump 
structure in $v_2(\eta)$ obtained in the standard version of the QGSM 
disappears.

\begin{figure}[ht]
\includegraphics[scale=0.525]{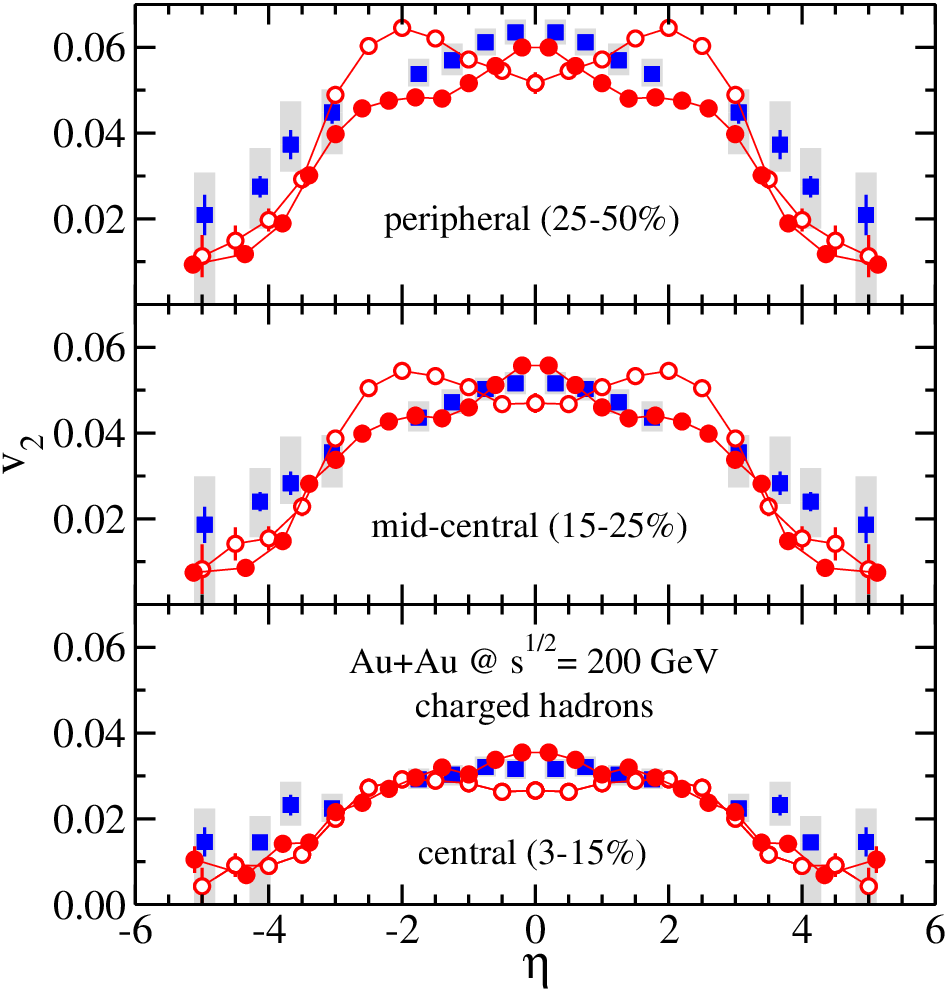}
\caption{Pseudorapidity distributions of $v_2$ for charged hadrons 
from $\sqrt{s_{NN}} = 200~{\rm GeV}$ Au+Au reactions for three centrality 
classes according to the PHOBOS analysis \cite{Back:2004mh}. 
The identification marking is the same as in Fig. \ref{fig:v2etahchar1}.
\label{fig:v2etahchar2}}
\end{figure}

This striking feature holds also for the centrality dependence of $v_2(\eta)$ 
as depicted in Fig. \ref{fig:v2etahchar2}. 
Here, the results from the QGSM simulations and the PHOBOS analysis 
(combined data from the hit- and track-based methods) for three different 
centrality classes are overlaid. The QGSM, including partonic recombination 
processes, is able to describe the magnitude as well as the pseudorapidity 
dependence of $v_2(\eta)$ remarkably well for all the centrality classes, 
ranging from central via midcentral to peripheral in accordance with the 
definitions in Ref. \cite{Back:2004mh}.

This is a highly non-trivial result, since a simultaneous description of 
both observables has neither been achieved by other standard string-cascade 
transport models such as relativistic quantum molecular dynamics (RQMD) or 
ultrarelativistic quantum molecular dynamics (UrQMD) 
\cite{Bleicher:2000sx,Lu:2006qn}, nor by purely hydrodynamical calculations 
\cite{Hirano:2001eu,Hirano:2002hv}. 
So far, only a hydrodynamics+cascade hybrid approach with Glauber model 
initial conditions was able to give a fair description of the experimental 
data, with the exception of the midrapidity region in the most central 
collision class \cite{Hirano:2005xf}. There it has been argued that the 
hadronic cascade provides the right amount of dissipation to bring the 
ideal fluid prediction down to the measured values, especially in peripheral 
collisions and away from midrapidity.

Following this argument, the $v_2$ results obtained with the extended QGSM 
may be interpreted in a complementary way as follows: In contrast to a highly 
dissipative hadronic medium, the parton recombination processes lead to a 
reduction of the mean free path in the very dense stages of a heavy-ion 
collision at midrapidity. Accordingly, the viscosity of this strongly 
interacting partonic medium is effectively lowered in comparison to the pure 
hadronic medium, i.e., the rearrangement processes on the partonic level 
reduce the amount of dissipation in the highly dense matter and enhance the 
elliptic flow, especially in the midrapidity region, to bring the theoretical 
predictions in line with the data. Thus far the QGSM upgraded by the locally 
density-dependent parton recombination mechanism quasi models the possible 
dynamics of a sQGP from a microscopical point of view.

\begin{figure}[b]
\includegraphics[scale=0.525]{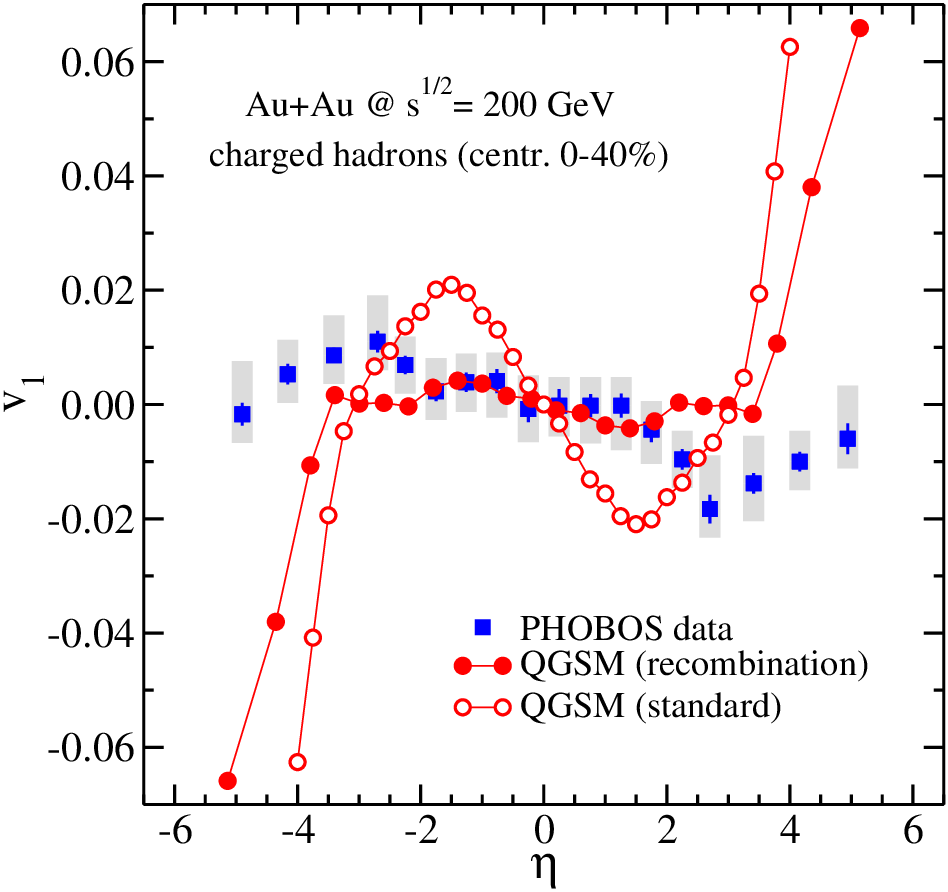}
\caption{Influence of parton recombination processes on the 
pseudorapidity shape of the directed flow $v_1$ for charged hadrons. 
The results obtained with the standard QGSM (open symbols) \cite{Burau:2004ev} 
and those from a simulation using the QGSM extended by parton recombination 
(filled symbols) are shown in comparison to PHOBOS data \cite{Back:2005pc}. 
The systematic errors of the experimental data are indicated by gray boxes 
and the statistical errors are indicated by bars.
\label{fig:v1etahchar}}
\end{figure}

A complementary observable regarding anisotropic flow phenomena is the 
directed flow $v_1$ of produced particles. In Fig. \ref{fig:v1etahchar}, 
the results for the pseudorapidity dependence of this flow component, 
i.e., $v_1(\eta)$ of charged final hadrons, obtained by QGSM simulations 
with and without parton recombination processes are compared to the 
corresponding experimental data from the PHOBOS Collaboration for 
0\% to 40\% central Au+Au collisions at the highest RHIC energy of 
$\sqrt{s_{NN}} = 200~{\rm GeV}$. 
The result of the standard QGSM shows a characteriatic \textsl{wiggle} 
structure extensively discussed in Ref. \cite{Burau:2004ev}, whereas the 
scenario with parton recombination is able to reproduce the experimentally 
observed directed flow very well in the broad midrapidity region where $v_1$ 
is essentially flat and close to zero. From the microscopical point of view, 
this is another hint for a strongly interacting partonic medium which exists 
during the early dense stage of ultrarelativistic heavy-ion collisions around 
midrapidity. Thus, our findings are in line with the results of an anisotropic 
flow study using a multiphase transport (AMPT) model that includes both 
initial partonic and final hadronic interactions \cite{Chen:2004vh}. 
There also the conclusion has been drawn that the matter produced during the 
early stage of Au+Au collisions at $\sqrt{s_{NN}} = 200~{\rm GeV}$ in the 
pseudorapidity region $|\eta| \le 3$ is dominated by partons.


\section{Summary and conclusions}
\label{sum}

In summary, it has been demonstrated that partonic fusion and recombination 
processes which occur in the very dense medium created in ultrarelativistic 
heavy-ion reactions during the early stages lead to short relaxation times 
and drive the system to fast kinetic equilibrium. The basis for these 
investigations was a microscopic transport model, namely the quark-gluon 
string model (QGSM) based on the color exchange mechanism for string 
formation, which has been extended by a locally density-dependent partonic 
recombination procedure to model effectively the dynamics of a strongly 
coupled quark plasma and final hadronic interactions.

Moreover, the pseudorapidity distributions of the anisotropy parameter 
$v_2(\eta)$ of final charged hadrons has been found to be intimately related 
to the corresponding dynamics. Fast equilibration due to parton recombination 
is necessary in order to obtain $v_2(\eta)$ profiles which are clearly peaked 
at midrapidity as seen in the data. We want to note that the other observables 
studied with the standard QGSM \cite{Burau:2004ev} turned out to be 
essentially robust against the inclusion of quark recombination. 
In particular the particle species dependence of $v_2(p_T)$ is still 
reproduced with the extended QGSM. The rapidity distribution for the final 
hadrons and their directed flow $v_1(\eta)$ are even considerably improved.

In conjunction with our aforementioned results for the local equilibration 
behavior, this is from the microscopical point of view a strong indication 
for the creation of a strongly interacting partonic medium in Au+Au collisions 
at RHIC that is thermally equilibrated on a very short time scale.

\begin{acknowledgments}
\label{acknowl}
The authors are grateful to L.V. Bravina and E.E. Zabrodin for fruitful 
discussions. This work was supported by the Bundesministerium f{\"u}r 
Bildung und Forschung (BMBF) under contract 06T\"U202.
\end{acknowledgments}


\end{document}